\documentclass[pdftex,twocolumn,epjc3]{svjour3}          

\RequirePackage[T1]{fontenc}

\smartqed  

\usepackage{graphicx}
\usepackage{flushend}
\usepackage[numbers,sort&compress]{natbib}
\usepackage[colorlinks,citecolor=blue,urlcolor=blue,linkcolor=blue]{hyperref}
\usepackage{widetext}
\usepackage{color}
\usepackage{dcolumn}
\usepackage{bm}
\usepackage{slashed}
\usepackage{amsmath}
\usepackage{latexsym}
\usepackage{amssymb}
\usepackage{mathrsfs}
\usepackage{amsfonts}
\usepackage{url}
\allowdisplaybreaks

\journalname{Eur. Phys. J. C}

\begin{document}

\title{Finite-distance gravitational deflection of massive particles by a rotating black hole in loop quantum gravity}


\author{Yang Huang\thanksref{addr1,addr2}
        \and
        Zhoujian Cao\thanksref{e1,addr1,addr2,addr3} 
}

\thankstext{e1}{corresponding author, zjcao@bnu.edu.cn}
\institute{Institute for Frontiers in Astronomy and Astrophysics, Beijing Normal University, Beijing 102206, China \label{addr1}
           \and
           Department of Astronomy, Beijing Normal University, Beijing 100875, China \label{addr2}
           \and
           School of Fundamental Physics and Mathematical Sciences, Hangzhou Institute for Advanced Study, UCAS, Hangzhou 310024, China \label{addr3}
}

\date{Received: date / Accepted: date}

\maketitle

\begin{abstract}
A rotating black hole in loop quantum gravity was constructed by Brahma, Chen, and Yeom based on a nonrotating counterpart using the revised Newman-Janis algorithm recently. For such spacetime, we investigate the weak gravitational deflection of massive particles to explore observational effects of the quantum correction. The purpose of this article is twofold. First, for Gibbons-Werner (GW) method, a geometric approach computing the deflection angle of particles in curved spacetimes, we refine its calculation and obtain a simplified formula. Second, by using GW method and our new formula, we work out the finite-distance weak deflection angle of massive particles for the rotating black hole in loop quantum gravity obtained by Brahma $et\ al.$ An analysis to our result reveals the repulsive effect of the quantum correction to particles. What's more, an observational constraint on the quantum parameter is obtained in solar system.
\end{abstract}
\maketitle

\section{Introduction}\label{sec-introduction}
The spacetime singularity at the beginning of the universe or inside a black hole is an inevitable issue for general relativity. It is believed that the quantum correction will be helpful for situations where classical general relativity breaks down. Although a well-built quantum gravity is still beyond our reach, many efforts aiming to resolve the spacetime singularity have been implemented. Loop quantum gravity (LQG), a background independent and nonperturbative approach to the unification of general relativity and quantum physics based on a quantum theory of geometry, has been one of the most potential candidates of quantum gravity theory. And the corresponding effective spacetime description developed in LQG is called polymer model.

In the standard model of cosmology, the singular big bang point serves as an origination of our universe
according to the singularity theorem proved by Hawking and Penrose \cite{penrose1965gravitational,hawking1967occurrence,hawking1970singularities}. For this singular point, small length scales and high curvatures are involved, thus the quantum effects can't be neglected. Quantum gravity, with its different dynamics on small scales, is expected to solve such problem. LQG predicts that the spatial geometry is discrete at Planck scale \cite{ashtekar2004background,rovelli2015covariant,han2007fundamental,thiemann2007modern}. By applying techniques in LQG to the homogeneous and isotropic geometry, people developed the loop quantum cosmology, in which a polymer scale is introduced to remove classical singularities \cite{bojowald2008loop,ashtekar2011loop,Agullo2016tjh}.

LQG also brings about a resolution of the singularity inside black holes. Inspired by LQG and considering the quantum corrections to gravity, one can modify the classical Hamiltonian and obtain a singularity-free solution that can be interpreted as a classical extension of spacetimes. Thus the singularity inside classical black holes is resolved by connecting the black hole region with the white hole region through a bounce. Specifically, a non-singular phase where the curvature and density are finite is introduced as a substitute for the singularity, such that infalling matter may bounce back and reemerge when the initial horizon evaporates. The resulting regular black hole is always called (effective) polymerized black hole and can be described by an explicit metric. Consequently, the well-developed mechanism for semiclassical black holes can be applied to polymerized black holes.
In the past years, a considerable number of polymerized black holes have been constructed, most of them are non-rotating models \cite{modesto2006loop,campiglia2007loop,bohmer2007loop,chiou2008phenomenological,modesto2008black,modesto2010semiclassical,caravelli2010spinning,corichi2016loop,perez2017quantum,olmedo2017black,ashtekar2018quantum,ashtekar2018quantumextension,bodendorfer2019effective,bodendorfer2019note,gambini2020spherically,kelly2020effective,bouhmadi2020asymptotic,kelly2020black,faraoni2020unsettling,gan2020properties,bojowald2020no,bodendorfer2021b,bodendorfer2021mass,sartini2021quantum,geiller2021symmetries}, and few are rotating \cite{caravelli2010spinning,liu2020shadow,brahma2021testing}. Many researches on various aspects of such effective geometries have also been carried out \cite{alesci2014particle,achour2018non,achour2018polymer,barrau2018status,moulin2019quantum,moulin2019overview,bouhmadi2020consistent,fu2022gravitational,barrau2011probing,gambini2014hawking,barrau2015black,ashtekar2020black,arbey2021hawking,arbey2021hawkingradiation,achour2020bouncing,achour2020bouncingcompact,achour2020towards,blanchette2021black,munch2021causal,liu2021extended,liu2022solar,yang2022destroying}. 
In this paper, we concentrate on the polymerized rotating black hole constructed by Brahma, Chen and Yeom \cite{brahma2021testing} (Brahma-Chen-Yeom polymerized rotating black hole), and investigate the weak gravitational deflection of particles by it.
It should be noted that the quantum correction is not the only scheme for regular black holes, there might be some regular black holes from vacuum bubbles \cite{brahma2020can,hwang2013firewall}, non-linear electrodynamics \cite{ayon1998regular,ayon1999new,bronnikov2000comment}, phantom matters and various modified gravity models \cite{nicolini2006noncommutative,frolov1990black}.

Since the observation of gravitational deflection of light passing near the Sun in 1919 \cite{dyson1920ix}, the gravitational lensing has been extensively studied in cosmology and astronomy \cite{bartelmann2010gravitational,wambsganss1998gravitational,blandford1992cosmological,dodelson2017gravitational}. Besides the photons, the massive particles such as neutrinos \cite{icecube2018neutrino}, massive gravitons \cite{abbott2017gw170814}, and cosmic rays \cite{letessier2011ultrahigh} are also significant messengers for exploring the universe.
The deflection angle is an important quantity in the research of the gravitational deflection of massless and massive particles, thus various approaches have been proposed to calculate it. In 2008, Gibbons and Werner introduced a geometric method to calculate the deflection angle of the photons in weak deflection limit for static spherically symmetric (SSS) and asymptotically flat spacetimes on the assumption that the source and the receiver are both at infinity \cite{gibbons2008applications}. It provides us a new geometric viewpoint to understand the gravitational deflection of light. For an SSS spacetime
\begin{equation}
    \mathrm{d}s^2=g_{tt}\mathrm{d}t^2+g_{rr}\mathrm{d}r^2+r^2(\mathrm{d}\theta^2+\sin^2\theta \mathrm{d}\phi^2),
\end{equation}
it can be proved, with Fermat's principle \cite{perlick2006fermat,perlick2000ray}, that the spatial projection of the light-like geodesic is the geodesic of the corresponding optical metric
\begin{equation}
    \mathrm{d}\tilde s^2=\frac{g_{rr}}{-g_{tt}}\mathrm{d}r^2+\frac{r^2}{-g_{tt}}(\mathrm{d}\theta^2+\sin^2\theta \mathrm{d}\phi^2).
\end{equation}
By applying the Gauss-Bonnet theorem to the optical space, Gibbons and Werner not only worked out the deflection angle, but also demonstrated that the gravitational lensing is a topological effect and the deflection angle is of global nature \cite{gibbons2009universal}.
Thereafter, a series of works utilizing GW method emerged. Werner extended this method to the stationary axially symmetric (SAS) spacetime by applying Naz{\i}m’s method to optical Randers-Finsler metric \cite{werner2012gravitational}. Crisnejo and Gallo extended GW method to the study of massive particles in SSS spacetimes with the help of Jacobi metric \cite{crisnejo2018weak}. Jusufi $et\ al.$ calculated the deflection angle of massive particles in SAS spacetimes \cite{jusufi2018gravitational,jusufi2019distinguishing}.

Considering the distance from the source and the receiver to a lens are both finite in reality, Ishihara $et\ al.$ defined a finite-distance deflection angle for photons in SSS spacetimes, and proved its geometric invariance with GW method \cite{ishihara2016gravitational,ishihara2017finite}. Ono $et/ al.$ derived the finite-distance deflection angle in an SAS spacetime with the generalized optical metric using GW method \cite{ono2017gravitomagnetic,ono2018deflection,ono2019deflection}. Furthermore, Li and Jia extended Ono's treatment to the massive particles in SAS spacetimes with the generalized Jacobi metric \cite{li2020thefinitedistance}. More related studies can be found in \cite{jusufi2016gravitational,jusufi2016light,jusufi2017quantum,jusufi2017deflection,jusufi2018deflection,sakalli2018analytical,jusufi2019light,haroon2019shadow,crisnejo2019higher,li2020finite,li2020gravitationaldeflection,li2020circular}. In fact, the work mentioned above shows that the finite-distance deflection angle introduced by Ishihara $et\ al.$ \cite{ishihara2016gravitational} can be regarded as a general definition in GW method. For the massless or massive particles in SSS or SAS spacetimes, one can calculate the finite-distance or infinite-distance deflection angle using this definition.

The contribution of this paper is twofold. (a) Through analyzing the double integral in the geometric expression of deflection angle in GW method, we refine the calculation and give a simplified formula, with which one can derive the result with fewer steps. (b) We obtain the finite-distance weak gravitational deflection angle of massive particles moving in the equatorial plane of Brahma-Chen-Yeom polymerized rotating black hole \cite{brahma2021testing} with GW method and our simplified formula. The result can be used to study the finite-distance or infinite-distance deflection angle of massless or massive particles in SSS or SAS counterparts. Moreover, with the observation data and the theoretical deflection angle of photons (reduced from that of massive particles), an observational constraint on the quantum parameter is given.

The remainder of this paper is organized as follows. In Sec.~\ref{sec-2}, the Brahma-Chen-Yeom polymerized rotating black hole is briefly introduced, and the motion of particles in such spacetime is investigated. In Sec.~\ref{sec-3}, we review the Gauss-Bonnet theorem and the Jacobi-Maupertuis Randers-Finsler (JMRF) metric, and show the application of GW method to an SAS spacetime. A simplified calculation formula of deflection angle in GW method is presented. In Sec.~\ref{sec-4}, the finite-distance weak gravitational deflection angle of massive particles in Brahma-Chen-Yeom polymerized rotating black is investigated, and an observational constraint on the quantum parameter is derived. Finally, we close our paper with a short conclusion. In the upcoming sections, we choose the spacetime signature ($-, +, +, +$) and geometric units such that the gravitational constant and the speed of light are set to one, $G=1$ and $c=1$.

\section{Brahma-Chen-Yeom polymerized rotating black} \label{sec-2}
\subsection{Basic introduction}
Taking the SSS black hole in LQG given by \cite{bodendorfer2021b,bodendorfer2021mass} as the seed metric, Brahma, Chen and Yeom obtained the rotating counterpart by using the revised Newman-Janis algorithm \cite{brahma2021testing},
\begin{equation}
    \begin{aligned}
        \mathrm{d}s^{2} = & -\left( 1-\frac{2M_0\mathcal{B}}{\rho ^{2}}\right) \mathrm{d}t^{2} -\frac{4aM_{0}\mathcal{B}\sin^{2} \theta }{\rho^2 } \mathrm{d}t \mathrm{d}\phi \\
        &+\rho ^{2} \mathrm{d}\theta ^{2} +\frac{\rho ^{2}}{\Delta } \mathrm{d}r^{2} +\frac{\Sigma \sin^{2} \theta }{\rho ^{2}} \mathrm{d}\phi ^{2},
        \end{aligned}
        \label{bhlqg}
\end{equation}
where
\begin{equation}
    \begin{aligned}
        & \Delta =8 \lambda M_B^2\mathcal{AB}^{2} +a^{2} ,\\
        & \Sigma =\left(\mathcal{B}^{2} +a^{2}\right)^{2} -a^{2} \Delta \sin^{2} \theta ,\\
        & M_0=\frac{1}{2}\mathcal{B}\left( 1-8 \lambda M_{B}^{2}\mathcal{A}\right) ,\\
        & \rho ^{2} =\mathcal{B}^{2} +a^{2}\cos^{2} \theta ,
        \end{aligned}
\end{equation}
with
\begin{equation}
\begin{aligned}
\mathcal{B}^{2} &=\frac{\lambda}{\sqrt{1+x^{2}}} \frac{M_B^{2}\left(x+\sqrt{1+x^{2}}\right)^{6}+M_W^{2}}{\left(x+\sqrt{1+x^{2}}\right)^{3}}, \\
\mathcal{A} &=\left(1-\frac{1}{\sqrt{2 A}} \frac{1}{\sqrt{1+x^{2}}}\right) \frac{1+x^{2}}{\mathcal{B}^{2}},
\end{aligned}
\end{equation}
and $x=r/(M_B\sqrt{8  \lambda})$, $M_B$ and $M_W$ are the mass of asymptotically Kerr black hole and white hole, respectively, $\lambda = \left[\lambda_k/(M_B M_W)\right]^{2/3}/2$ is a dimensionless and nonnegative parameter, $\lambda_k$ is a quantum parameter originated from holonomy modifications \cite{bodendorfer2021b}.

The metric \eqref{bhlqg} is defined in terms of the radial variable $r\in \left(-\infty,\infty\right)$, 
and we only focus on the most interesting and meaningful case $M_B=M_W$, namely the spacetime that is symmetric with respect to the transition surface ($r=0$) \cite{yang2022destroying}. This polymerized black hole is nonsingular everywhere and reduces to Kerr black hole asymptotically. When $a=0$, it reduces to the nonrotating counterpart \cite{bodendorfer2021mass,bodendorfer2021b}. When $\lambda=0$, it reduces to the Kerr spacetime. When $a=0$ and $\lambda=0$, it reduces to the Schwarzschild spacetime. What's more, when $a=0$ and $M_B=M_W=0$, it reduces to a flat spacetime \cite{gan2020properties}, this behaviour satisfies an essential consistency check which some quantum gravity inspired solutions do not meet \cite{hossenfelder2010model}. See \cite{brahma2021testing} to get more details about Brahma-Chen-Yeom polymerized rotating black hole.

\subsection{Motion of massive particles}
Consider an SAS spacetime whose metric reads
\begin{equation}
    \begin{aligned}
    \mathrm{d}s^2= & g_{\mu\nu}(r,\theta)\mathrm{d}x^\mu \mathrm{d}x^\nu \\
        = & g_{tt} \mathrm{d} t^2 +2g_{t\phi}\mathrm{d}t\mathrm{d}\phi+ g_{rr} \mathrm{d}r^2+ g_{\theta\theta} \mathrm{d}\theta^2+g_{\phi\phi}\mathrm{d}\phi^2,
\end{aligned}
\end{equation}
we focus on the motion of massive particles in the equatorial plane, thus the unit speed condition yields
\begin{equation}
    g_{tt}\dot{t}^2+2g_{t\phi}\dot{t}\dot{\phi}+g_{rr}\dot{r}^2+g_{\phi\phi}\dot{\phi}^2=-1,
    \label{unitcondition}
\end{equation}
in which the dot means the derivative with respect to the affine parameter. Additionally, based on two Killing vectors $(\partial/\partial t)^a$ and $(\partial/\partial \phi)^a$, we have two conserved quantities
\begin{equation}
        E  = -m\left( g_{tt}\dot{t}+g_{t\phi} \dot{\phi} \right), \quad
        L  = m\left(g_{t\phi} \dot{t}+g_{\phi\phi}\dot{\phi}\right),
    \label{ELtphi}
\end{equation}
which are energy and angular momentum of particles, respectively. Let $v$ denote the velocity of the particles at infinity, and $b$ denote the impact parameter, then we have
\begin{equation}
        E = \frac{m}{\sqrt{1-v^2}}, \qquad       L = \frac{mbv}{\sqrt{1-v^2}}.
    \label{ELvb}
\end{equation}
Combining Eqs.~\eqref{unitcondition}, \eqref{ELtphi} and \eqref{ELvb} yields
\begin{equation}
    \begin{aligned}
        \left(\frac{\mathrm{d}u}{\mathrm{d}\phi} \right)^2 = &    \frac{u^4 \left(g_{t\phi}^2-g_{tt} g_{\phi \phi}\right) }{g_{rr} ( g_{tt}b v +g_{t\phi})^2}
    \left[  g_{tt} b^2 v^2+2  g_{t\phi}bv + \right.  \\
    & \left. g_{\phi \phi } \left(1+g_{tt}-g_{tt} v^2\right)+g_{t\phi}^2 \left(v^2-1\right)\right],
\end{aligned}
\label{dudphi}
\end{equation}
in which $u= 1/r$.

Substituting metric \eqref{bhlqg} into Eq.~\eqref{dudphi}, we obtain the orbit of free massive particles in the equatorial plane of Brahma-Chen-Yeom polymerized rotating black hole with iterative method
\begin{equation}
    \begin{aligned}
        u & =\frac{\sin \phi }{b} +M\frac{1+v^{2}\cos^{2} \phi }{b^{2} v^{2}} +M^{2}\frac{\cos \phi }{8b^{3} v^{4}}\bigg\{2v^{2} \phi \cdot \\
         & \left[ 4\lambda \left( v^{2} +3\right) -3\left( v^{2} +4\right)\right] -(8\lambda +3)v^{4}\sin (2\phi )\\
         & +4\left[ 2\lambda v^{4} +(4-6\lambda )v^{2} +1\right]\tan \phi \bigg\} -Ma\frac{2}{b^{3} v}\\
         & +\mathcal{O} (M^{3} , M^2a, Ma^{2}, a^3 ).
        \end{aligned}
    \label{uofphi}
\end{equation}
According to the above formula, we can also obtain the iterative solution of $\phi$ for the orbit
\begin{equation}
    \phi = \begin{cases}
    \Phi(u), & \text{if  } \left| \phi \right| <\frac{\pi}{2}, \\
    \pi - \Phi(u) , & \text{if  } \left| \phi \right| >\frac{\pi}{2},
        \end{cases}
        \label{phigamma}
\end{equation}
where
\begin{equation}
    \begin{aligned}
        \Phi (u)= & \arcsin (bu)+M\frac{b^{2} v^{2} u^{2} -v^{2} -1}{bv^{2}\sqrt{1-b^{2} u^{2}}} +M^{2}\frac{1}{4b^{2} v^{4}}\Bigg\{\\
         & \frac{bu\left[ 12(\lambda -1)v^{2} -v^{4}\left( (8\lambda -3)b^{2} u^{2} -4\lambda +3\right)\right]}{\sqrt{1-b^{2} u^{2}}}\\
         & +v^{2}\left[ 3\left( v^{2} +4\right) -4\lambda \left( v^{2} +3\right)\right]\arcsin (bu)\\
         & +\frac{2b^{3} u^{3}}{\left( 1-b^{2} u^{2}\right)^{3/2}}\Bigg\}+Ma\frac{2}{b^{2} v\sqrt{1-b^{2} u^{2}}}\\
         &  -a^{2}\frac{bu^{2}}{2\sqrt{1-b^{2} u^{2}}}  +\mathcal{O} (M^{3} ,M^{2} a ,Ma^{2} ,a^{3} ).
        \end{aligned}
\end{equation}

\section{Refine the calculation of GW method} \label{sec-3}
We focus on the weak gravitational deflection of massive particles in the equatorial plane of SAS spacetimes. Firstly, an introduction to the Gauss-Bonnet theorem and the JMRF metric is given. Then, we show how to apply GW method to SAS spacetimes and obtain the geometric formula of the deflection angle. Finally, we refine the calculation of the geometric formula and obtain a simplified formula for the deflection angle.

\subsection{Gauss-Bonnet theorem and JMRF metric}
Given a compact two-dimensional Riemannian manifold $D$ with piecewise smooth boundary $\partial D=\cup_i \partial D_i (i=1,2,\cdots)$. $K$ denotes the Gaussian curvature of $D$, $\kappa$ denotes the geodesic curvature of $\partial D_i$, $\eta_i$ is the jump angle (or exterior angle) at the $i$-th vertex in the sense of positive. Then the Gauss-Bonnet theorem states \cite{manfredo1976carmo}
\begin{equation}
    \iint_D K \mathrm{d}S + \sum_i \int_{\partial D_i} \kappa \mathrm{d}l+\sum_i \eta_i = 2\pi \chi(D),
\end{equation}
where $\mathrm{d}S$ and $\mathrm{d}l$ are the area element and line element, respectively. $\chi(D)$ is the Euler characteristic number of $D$. Since we only concern the motion in the equatorial plane, the most basic version of the Gauss-Bonnet theorem is sufficient for us. More generalizations of Gauss-Bonnet theorem to various situations can be found in \cite{sampson1973theorem,atiyah1959riemann,getzler1986short,walter1975generalized}.

JMRF metric is constructed by Chanda $et\ al.$ based on the principle of least action of Maupertuis \cite{chanda2019jacobi}.
Consider a coordinate $system$ $(t,\textbf{x})$, $\left(\partial/\partial t\right)^a$ is the time-like Killing vector related to the stationarity of the spacetime, then the metric of the stationary spacetime can be expressed as
\begin{equation}
    \mathrm{d}s^2=g_{\mu\nu} (\textbf{x}) \mathrm{d}x^\mu \mathrm{d}x^\nu.
\end{equation}
The spatial projection of a light-like or time-like geodesic in the stationary spacetime can be described as a geodesic in the corresponding JMRF space equiped with the Randers-Finsler metric \cite{chanda2019jacobi}
\begin{equation}
    \mathrm{d}\tilde{s} = \sqrt{\alpha_{ij} \mathrm{d}x^i \mathrm{d}x^j} + \beta_i \mathrm{d}x^i,
    \label{JMRF}
\end{equation}
in which
\begin{align}
    \alpha_{ij} &=  \frac{E^2 +m^2 g_{00}}{-g_{00}} \left(g_{ij}-\frac{g_{0i}g_{0j}}{g_{00}}\right),  \label{alphaij} \\
    \beta_i & = -E \frac{g_{0i}}{g_{00}} \label{betai}.
\end{align}
$E$ is the conserved quantity corresponding to the Killing vector $\left(\partial/\partial t\right)^a$, $m$ is the rest mass of the particle. For simplicity, the three-dimensional space determined by Riemannian metric $\alpha_{ij}$ is denoted as $M^{(\alpha 3)}$. What's more, a geodesic in JMRF space can be put in one-to-one correspondence with a curve, denoted by $\gamma$, in $M^{(\alpha 3)}$ \cite{li2020thefinitedistance}. The deviation between $\gamma$ and the geodesic in $M^{(\alpha 3)}$ is described by the one-form $\beta_i$. Apparently, for a static spacetime, $\beta_i=0$, and $\gamma$ is a geodesic in $M^{(\alpha 3)}$.

\subsection{Apply GW method to SAS spacetimes}
\begin{figure}[!ht]
    \includegraphics[width=0.9\columnwidth]{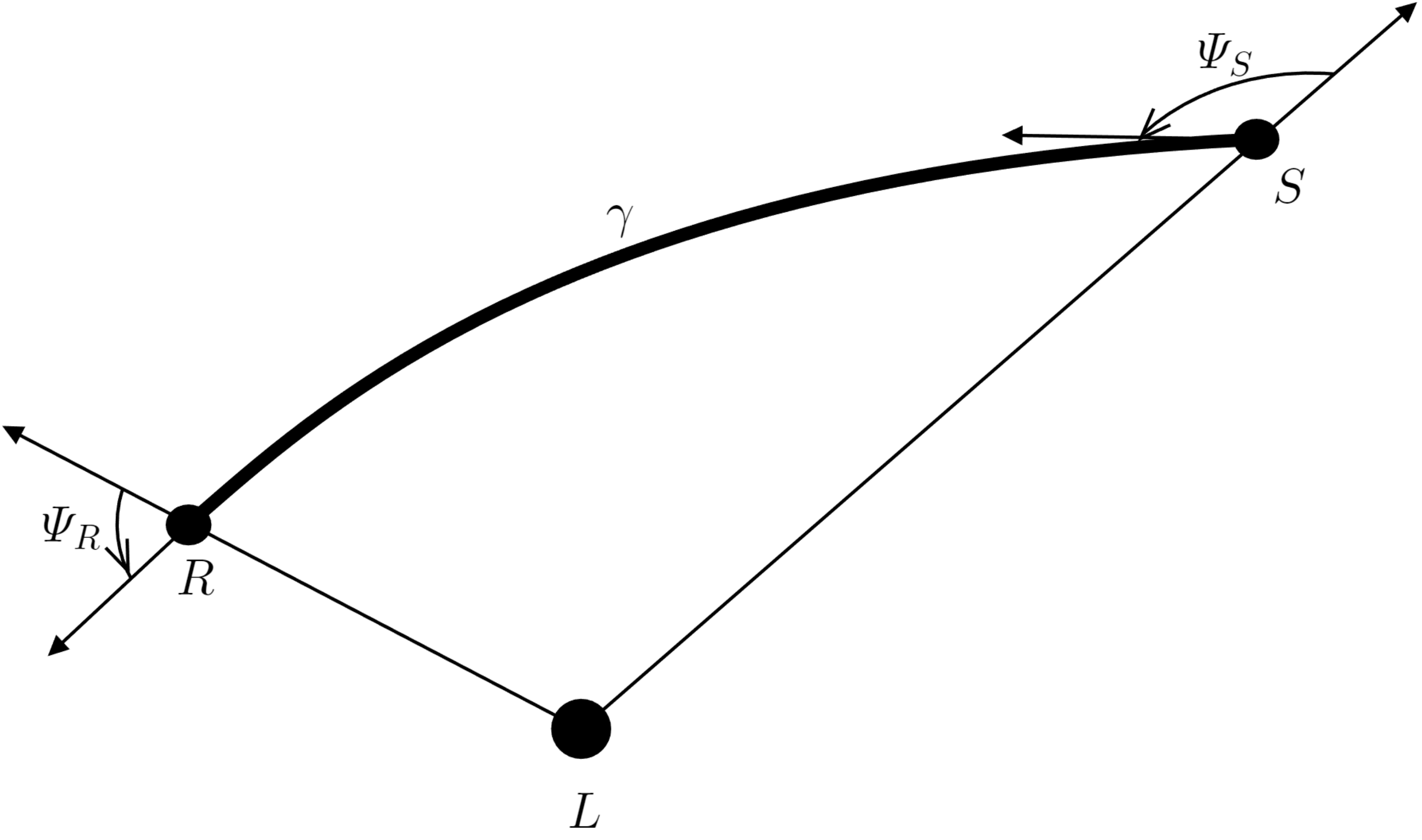}
    \caption{$\overset{\curvearrowright}{S R}$ is the trajectory of a particle in $M^{(\alpha 2)}$. The finite-distance deflection angle is defined as $\delta = \Psi_R - \Psi_S + \phi_R - \phi_S$.}
    \label{fig-1}
\end{figure}

For an SAS asymptotically flat spacetime, when $\theta=\pi/2$, the corresponding $M^{(\alpha 3)}$ reduces to a two-dimensional Riemannian space which is dentoed by $M^{(\alpha 2)}$. The metric of $M^{(\alpha 2)}$ can be expressed as
\begin{equation}
    \mathrm{d}l^2=\alpha_{rr}(r) \mathrm{d}r^2+\alpha_{\phi\phi}(r)\mathrm{d}\phi^2
    \label{metricalpha}
\end{equation}
in the orthogonal coordinate system. As shown in Fig.~\ref{fig-1}, in the space $M^{(\alpha 2)}$, $S$ and $R$ represent the source and the receiver of a free particle, respectively. The curve $\gamma = \overset{\curvearrowright}{ S R}$ corresponds to the spatial projection of the orbit. $L$ represents the lens. $\Psi_R$ and $\Psi_S$ are angles between the outward radial vector and the tangent along $\gamma$ at $R$ and $S$, respectively. Then the finite-distance deflection angle defined by Ishihara $et\ al.$ reads \cite{ishihara2016gravitational}
\begin{equation}
    \delta = \Psi_R - \Psi_S + \phi_R - \phi_S,
    \label{angledef}
\end{equation}
in which $\phi_R$ and $\phi_S$ are the azimuthal coordinate (associated with the rotational Killing vector in original four-dimensional spacetimes) of $R$ and $S$, respectively.

\begin{figure}[!ht]
    \includegraphics[width=0.9\columnwidth]{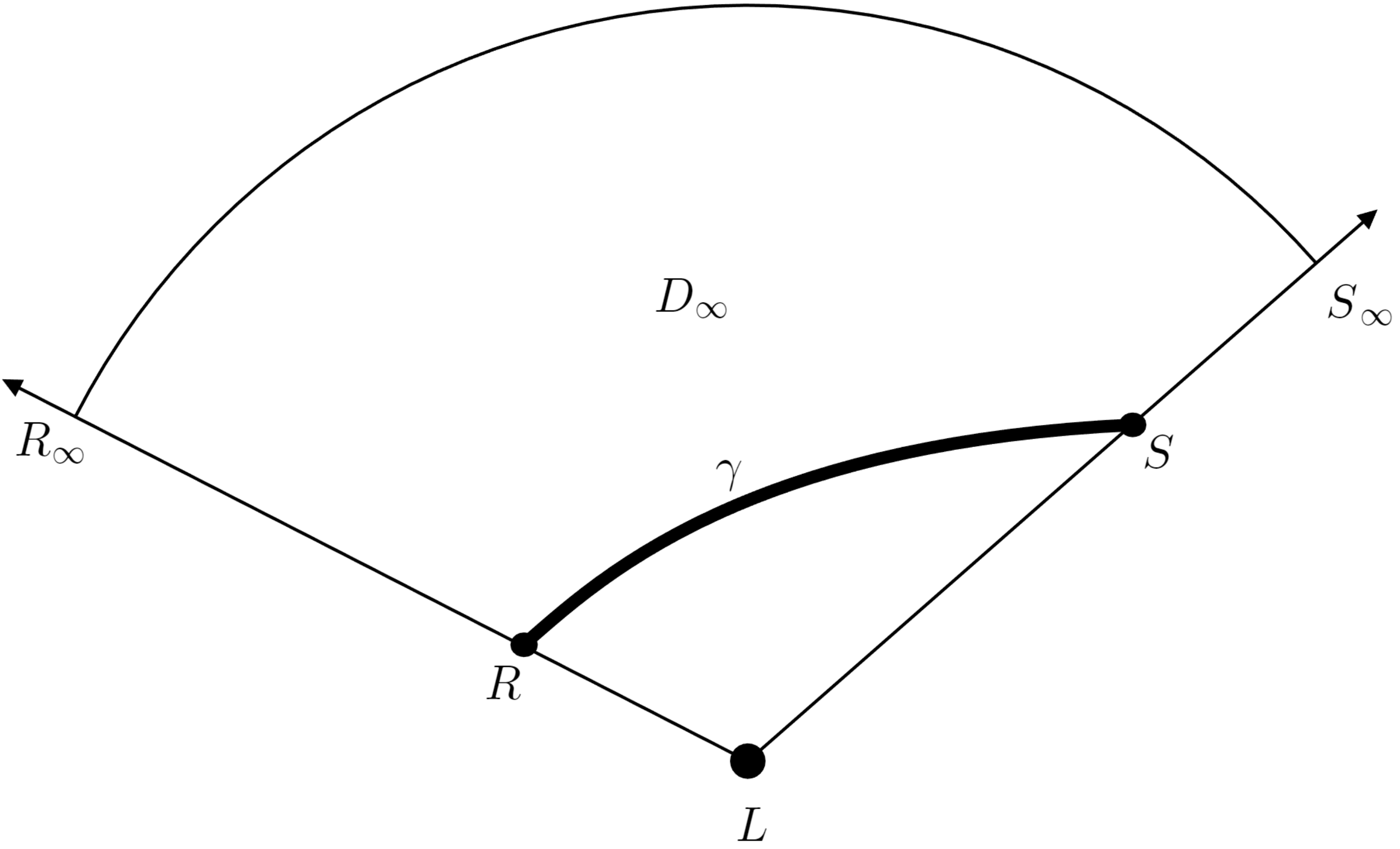}
    \caption{Sketch of the region $D_\infty = ^{R_\infty}_{R}\square_{S}^{S_\infty}$.}
    \label{fig-2}
\end{figure}

We consider the weak gravitational deflection of the unbound free particles. As shown in Fig.~\ref{fig-2}, the circular arc segment $\overset{\curvearrowright}{S_\infty R_\infty}$ with $r = \infty$ centered at $L$ intersects with two outgoing radial geodesics $\overrightarrow{SS}_\infty$ and $\overrightarrow{RR}_\infty$ at $S_\infty$ and $R_\infty$, respectively. Thus we can construct a region $D_\infty = ^{R_\infty}_{R}\square_{S}^{S_\infty}$.

The geodesic curvature of $\overrightarrow{SS}_\infty$ vanishes (the proof is shown in \ref{appendix-1}), so does $\overrightarrow{RR}_\infty$. The jump angles at $S_\infty$ and $R_\infty$ are $\pi/2$. The Euler characteristic number $\chi(D_\infty)$ is one since the nonsingularity of $D_\infty$. Applying the Gauss-Bonnet theorem to $D_\infty$ yields
\begin{equation}
    \iint_{D_\infty}  K \mathrm{d}S + \int_{\overset{\curvearrowright}{RS}} \kappa \mathrm{d}l + \int_{\overset{\curvearrowright}{S_\infty R_\infty}} \kappa \mathrm{d}l  +\Psi_R - \Psi_S =0 \label{GBTinDinfty}.
\end{equation}
On account of the asymptotics of the $M^{(\alpha 2)}$, for the infinite circular arc segment, we have $\mathrm{d}l=r_\infty \mathrm{d}\phi$, where $r_\infty$ represents the radial coordinate of $\overset{\curvearrowright}{S_\infty R_\infty}$. The geodesic curvature becomes the ordinary curvature in the asymptotically flat region, namely $\kappa=1/r_\infty$. Then
\begin{equation}
\int_{\overset{\curvearrowright}{S_\infty R_\infty}} \kappa \mathrm{d}l
    = \int_{\phi_S}^{\phi_R} \frac{1}{r_\infty} r_\infty \mathrm{d}\phi = \phi_R- \phi_S.
    \label{cediqulvinfty}
\end{equation}
Substituting Eq.~\eqref{cediqulvinfty} into Eq.~\eqref{GBTinDinfty}, we obtain
\begin{equation}
    \delta = - \iint_{D_\infty} K \mathrm{d}S   + \int_{\gamma} \kappa \mathrm{d}l,
    \label{geometricproof}
\end{equation}
which indicates that the gravitational deflection angle can be expressed by geometric quantities, namely, is geometric invariant and can be seen as a global effect.

In previous literatures, people routinely obtain the deflection angle by directly calculate all integrals in Eq.~\eqref{geometricproof}. Such treatment is complicated especially for the complex surface integral. One can see the usual calculation process of Eq.~\eqref{geometricproof} for photons in \cite{ono2017gravitomagnetic}, and massive particles in \cite{li2020thefinitedistance}.

\subsection{Refinement for the calculation of GW method}
In previous works, authors substitute a specific metric into the geometric expression Eq.~\eqref{geometricproof} without any pre-simplification to the surface integral of Gaussian curvature $\iint_{D_\infty}K \mathrm{d}S$. This means they have to face the complicated surface integral. Starting from a specific metric at hand, they work out the Gaussian curvature of $M^{(\alpha 2)}$ with \cite{werner2012gravitational}
\begin{equation}
    K=\frac{1}{\sqrt{\alpha}}\left[\frac{\partial}{\partial \phi}(\frac{\sqrt{\alpha}}{\alpha_{rr}} \Gamma^{\phi}_{rr})-\frac{\partial}{\partial r}(\frac{ \sqrt{\alpha}}{\alpha_{rr}} \Gamma^{\phi}_{r\phi}) \right],
    \label{gaosiqulv}
\end{equation}
in the first step. $\alpha$ is the determinant of metric~\eqref{metricalpha}, $\Gamma$ denotes the Christoffel symbol with respect to $M^{(\alpha 2)}$. Then the indefinite integral of the integrand $K\sqrt{\alpha}$ with respect to the radial coordinate is performed. Next, they obtain the definite integral with the integration range $\left(r_\gamma, \infty\right)$, where $r_\gamma$ denotes the radial coordinate of $\gamma$. Finally, the calculation ends with the definite integral with respect to the azimuthal coordinate.

Carefully analyzing the double integral before calculating it with specific metrics, we find that the first three steps of routine calculation can be replaced by a simple formula. Since the SAS metric does not contain the azimuthal coordinate, by using metric \eqref{metricalpha} and Eq.~\eqref{gaosiqulv}, we have
\begin{equation}
    \begin{aligned}
        \int K\sqrt{\alpha} \mathrm{d}r = &\int \left [ \frac{\partial}{\partial \phi}(\frac{\sqrt{\alpha}}{\alpha_{rr}} \Gamma^{\phi}_{rr})-\frac{\partial}{\partial r}(\frac{ \sqrt{\alpha}}{\alpha_{rr}} \Gamma^{\phi}_{r\phi}) \right] \mathrm{d}r \\
            = & \int -\frac{\partial}{\partial r}(\frac{ \sqrt{\alpha}}{\alpha_{rr}} \Gamma^{\phi}_{r\phi}) \mathrm{d}r \\
            = &- \frac{\alpha_{\phi\phi,r}}{2\sqrt{\alpha}} + Const .
    \end{aligned}
    \label{rint}
    \end{equation}
As a result,
\begin{equation}
    \begin{aligned}
        \iint_{D_\infty} K \mathrm{d}S  = & \int_{\phi_S}^{\phi_R} \int_{r_\gamma}^{\infty} K \sqrt{\alpha} \mathrm{d}r \mathrm{d}\phi     \\
        =& \int_{\phi_S}^{\phi_R} \left[ H(\infty)-H(r_\gamma) \right] \mathrm{d}\phi,
    \end{aligned}
    \label{Dgammagaosi}
\end{equation}
in which we have introduced a notation according to Eq.~\eqref{rint}
\begin{equation}
    H(r) = - \frac{\alpha_{\phi\phi,r}}{2\sqrt{\alpha}}.
    \label{Hr}
\end{equation}
On account of the asymptotics of $M^{(\alpha 2)}$, we have
\begin{equation}
    H(\infty)=-1.
    \label{Hinfty}
\end{equation}
With Eqs.~\eqref{Dgammagaosi} and \eqref{Hinfty}, the geometric expression Eq.~\eqref{geometricproof} can be rewritten as
\begin{equation}
    \delta = \int_{\phi_S}^{\phi_R} \left[1+ H(r_\gamma)\right] \mathrm{d}\phi + \int_{\gamma} \kappa \mathrm{d}l \label{delta2},
\end{equation}
which can be regarded as a simplified calculation formula of the finite-distance deflection angle. With Eq.~\eqref{delta2}, the double integral can be replaced by a single integral of a simple integrand. That is to say, one can work out the result more easily with fewer and simpler calculation steps. Such simplification can facilitate the calculation of integral of Gaussian curvature in actual implementation of GW method, especially for the highly accurate results and complicated spacetimes. 

On the one hand, Eq.\eqref{geometricproof} shows that the gravitational deflection can be regarded as a global effect. On the other hand, Eq.\eqref{delta2} only involves the quantities along the trajectory, which is also consistent with our empirical judgment that the deflection angle should be completely determined by the trajectory itself.

\section{Finite-distance deflection angle of massive particles for a rotating black hole in LQG} 
\label{sec-4}
\subsection{Finite-distance deflection angle}
In this subsection, we calculate the finite-distance weak gravitational deflection angle of massive particles for Brahma-Chen-Yeom polymerized rotating black hole with Eq.~\eqref{delta2}.

We first calculate $\int_{\phi_S}^{\phi_R} [1+ H(r_\gamma)] \mathrm{d}\phi$. Substituting Eq.~\eqref{bhlqg} into Eq.~\eqref{alphaij}, and setting $\theta=\pi/2$, the metric of the corresponding $M^{(\alpha 2)}$ becomes
\begin{equation}
    \mathrm{d} l^2 = \alpha_{rr} \mathrm{d}r^2 + \alpha_{\phi\phi} \mathrm{d} \phi^2,
    \label{M2LQG}
\end{equation}
where
\begin{widetext}
\begin{equation}
    \begin{aligned}
        \alpha _{rr} = & \frac{\left( 2\lambda M^{2} +r^{2}\right)\left[ 16\lambda ^{2} M^{4} +2\lambda M^{2} r\left( 3\sqrt{8\lambda M^{2} +r^{2}} +5r\right) +r^{3}\left(\sqrt{8\lambda M^{2} +r^{2}} +r\right)\right]}{a^{2}\sqrt{8\lambda M^{2} +r^{2}} +8\lambda M^{2}\left(\sqrt{8\lambda M^{2} +r^{2}} -2M\right) +r^{2}\left(\sqrt{8\lambda M^{2} +r^{2}} -2M\right)}\\
            & \cdot \frac{4\left[ 2Mm^{2}\sqrt{8\lambda M^{2} +r^{2}} +2\lambda M^{2}\left( E^{2} -m^{2}\right) +r^{2}\left( E^{2} -m^{2}\right)\right]}{\left(\sqrt{8\lambda M^{2} +r^{2}} +r\right)^{3}\left( -2M\sqrt{8\lambda M^{2} +r^{2}} +8\lambda M^{2} +r^{2}\right)} ,\\
            & \\
        \alpha _{\phi \phi } = & \frac{m^{2}\left( 2\lambda M^{2} +r^{2}\right)\left[ 2Mm^{2}\sqrt{8\lambda M^{2} +r^{2}} +2\lambda M^{2}\left( E^{2} -4m^{2}\right) +r^{2}\left( E^{2} -m^{2}\right)\right]}{\left( -2M\sqrt{8\lambda M^{2} +r^{2}} +8\lambda M^{2} +r^{2}\right)^{2}} \\
        &\cdot \left( a^{2} -2M\sqrt{8\lambda M^{2} +r^{2}} +8\lambda M^{2} +r^{2}\right) .
        \end{aligned}
\end{equation}
Then substituting Eq.~\eqref{M2LQG} into Eq.~\eqref{Hr}, we get
\begin{equation}
    H(r)=-1+M\frac{v^2+1 }{r v^2}+M^2\frac{ \left(1-4 \lambda\right) v^4+\left(6-12 \lambda\right) v^2-4}{2 r^2 v^4}+\mathcal{O}({M^3,M^2 a, Ma^2, a^3}).
\end{equation}
Thus
\begin{equation}
\begin{aligned}
        \int _{\phi _{S}}^{\phi _{R}} \left[ 1+H(r_{\gamma } )\right]\mathrm{d} \phi =& M\frac{\left( v^{2} +1\right)\left(\sqrt{1-b^{2} u_R^2} +\sqrt{1-b^{2} u_S^2}\right)}{bv^{2}} \\
        & + M^2\left\{
            \frac{b^{2} u_R^3\left[ \lambda \left( 4v^{4} +12v^{2}\right) -3v^{4} -8v^{2} +4\right] +u_R v^{2}\left[ -4\lambda \left( v^{2} +3\right) +3v^{2} +12\right]}{4bv^{4}\sqrt{1-b^{2} u_R^2}} \right. \\
        & +\frac{b^{2} u_{S}^{3}\left[ \lambda \left( 4v^{4} +12v^{2}\right) -3v^{4} -8v^{2} +4\right] +u_{S} v^{2}\left[ -4\lambda \left( v^{2} +3\right) +3v^{2} +12\right]}{4bv^{4}\sqrt{1-b^{2} u_S^2}} \\
        & +\left. \frac{\left[ 4\lambda \left( v^{2} +3\right) -3v^{2} -12\right][\arcsin (bu_{R} )+\arcsin (bu_{S} )-\pi ]}{4b^{2} v^{2}}\right\} \\
        & +\mathcal{O}({M^3,M^2 a, Ma^2, a^3}),
        \label{deltaA}
\end{aligned}
\end{equation}
where we have used the trajectory Eq.~\eqref{uofphi}, $\phi_S=\Phi(u_S)$ and $\phi_R=\pi-\Phi(u_R)$.

Next, we calculate $\int_{\gamma} \kappa \mathrm{d}l$. The geodesic curvature of $\gamma$ in $M^{(\alpha 2)}$ states \cite{ono2017gravitomagnetic}
\begin{equation}
    \kappa_\gamma=-\frac{\beta_{\phi, r}}{\sqrt{\alpha \cdot \alpha^{\theta \theta}}} = -\frac{2 Ma \sqrt{1-v^2} \sin ^3 \phi}{b^3 m v^2}+\mathcal{O}({M^3,M^2 a, Ma^2, a^3}).
\end{equation}
Since $\kappa_\gamma$ does not contain the zeroth-order and first-order terms, we can only retain the zeroth-order term of the line element along the trajectory $\gamma$
\begin{equation}
    \mathrm{d}l=\frac{mbv}{\sin^2\phi \sqrt{1-v^2}} \mathrm{d}\phi + \mathcal{O}(M,a).
\end{equation}
Then
\begin{equation}
    \begin{aligned}
    \int_{\gamma} \kappa \mathrm{d}l
    = & -\frac{2 Ma}{b^{2} v} \int_{\phi_{S}}^{\phi_{R}} \sin \phi \mathrm{d} \phi+\mathcal{O}\left(M^{3}, M^{2} a, a^{2} M, a^{3}\right) \\
    = & -\frac{2 Ma \left(\sqrt{1-b^2 u_R^2}+\sqrt{1-b^2 u_S^2}\right)}{b^2 v}+\mathcal{O}(M^3, M^2a, a^2M, a^3),
    \label{deltaB}
\end{aligned}
\end{equation}
in which we have used the trajectory \eqref{uofphi}, $\phi_S=\Phi(u_S)$ and $\phi_R=\pi-\Phi(u_R)$.

Finally, substituting Eqs.~\eqref{deltaA} and \eqref{deltaB} into Eq.~\eqref{delta2}, we obtain the finite-distance deflection angle
\begin{equation}
    \begin{aligned}
        \delta = & M\frac{\left( v^{2} +1\right)\left(\sqrt{1-b^{2} u_{R}^{2}} +\sqrt{1-b^{2} u_{S}^{2}}\right)}{bv^{2}} +\\
            & M^{2}\left\{\frac{b^{2} u_{R}^{3}\left[ \lambda \left( 4v^{4} +12v^{2}\right) -3v^{4} -8v^{2} +4\right] +v^{2} u_{R}\left[ -4\lambda \left( v^{2} +3\right) +3v^{2} +12\right]}{4bv^{4}\sqrt{1-b^{2} u_{R}^{2}}}\right. \\
            & +\frac{b^{2} u_{S}^{3}\left[ \lambda \left( 4v^{4} +12v^{2}\right) -3v^{4} -8v^{2} +4\right] +v^{2} u_{S}\left[ -4\lambda \left( v^{2} +3\right) +3v^{2} +12\right]}{4bv^{4}\sqrt{1-b^{2} u_{S}^{2}}}\\
            & \left. +\frac{\left[ 4\lambda \left( v^{2} +3\right) -3v^{2} -12\right][\arcsin (bu_{R} )+\arcsin (bu_{S} )-\pi ]}{4b^{2} v^{2}} \right\} \\
            & -Ma\frac{2\left(\sqrt{1-b^{2} u_{R}^{2}} +\sqrt{1-b^{2} u_{S}^{2}}\right)}{b^{2} v} +\mathcal{O}(M^3, M^2a, a^2M, a^3) .
        \end{aligned}
        \label{deltalqg}
\end{equation}
\end{widetext}

\subsection{Some discussion about the result}

If the spin parameter $a$ is positive, the above result corresponds to the prograde orbit where the direction of motion of particles is the same as that of the spin of black holes. Conversely, if $a$ is negative, the orbit is retrograde. Additionally, exchanging $u_R$ and $u_S$ will not affect this result, since the orbit is reversible.
The quantum parameter $\lambda$ appears in the second order. We rearrange Eq.~\eqref{deltalqg} and recast all terms containing $\lambda$ as
\begin{equation}
    \begin{aligned}
        &-\frac{\lambda M^2 \left(v^2+3\right)}{b^2 v^2} \bigg[\arccos\left(b u_R\right)+\arccos\left(b u_S\right)\\
        & +b u_R \sqrt{1-b^2 u_R^2}+b u_S \sqrt{1-b^2 u_S^2}   \bigg].
        \label{lambdaTerm}
    \end{aligned}
\end{equation}
Obviously the deflection angle depends linearly on the $\lambda$, and the negative-definite coefficient indicates that the quantum correction provides a repulsive effect to particles, which is contrary to the attractive effect of the mass of black hole.

\begin{figure}[!ht]
    \centering
    \includegraphics[width=1\columnwidth]{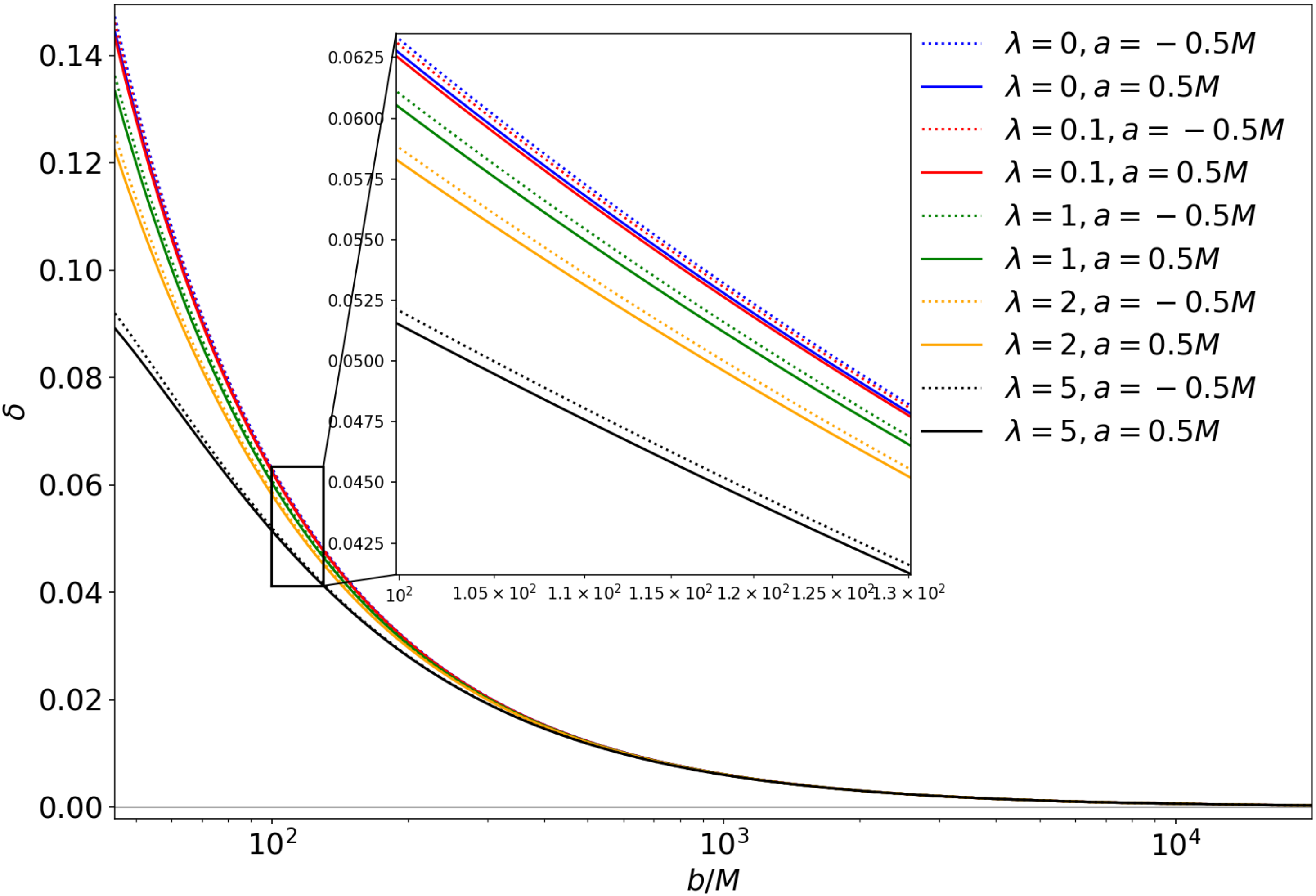}
    \caption{The finite-distance weak gravitational deflection angle of massive particles in Brahma-Chen-Yeom rotating black hole. $r_S=r_R=10^8M$, $v=0.7c$. We choose a large initial impact parameter since Eq.~\eqref{deltalqg} is a second-order post-Newtonian result for the weak gravitational lensing.}
    \label{fig-3}
\end{figure}

\begin{figure}[!ht]
    \centering
    \includegraphics[width=1\columnwidth]{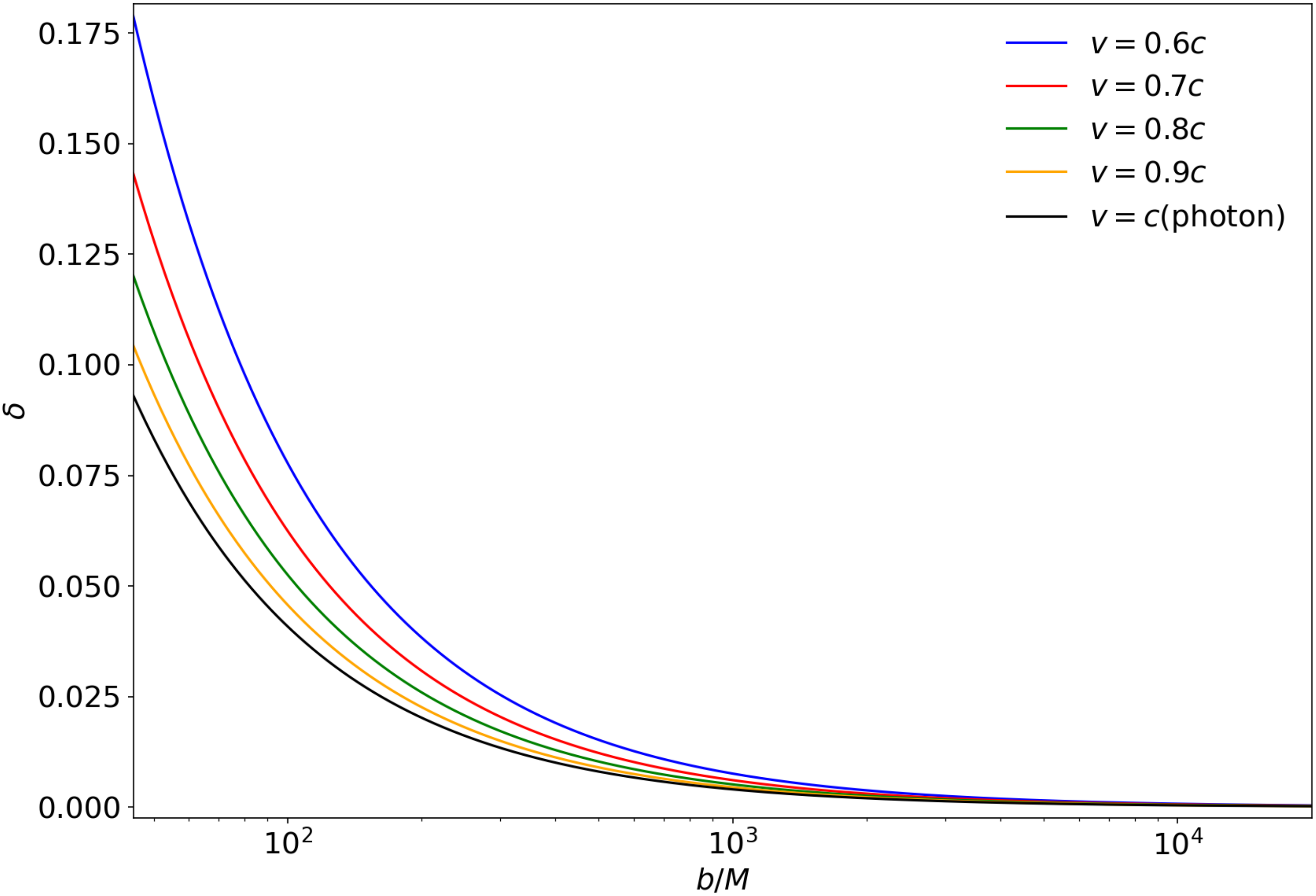}
    \caption{The finite-distance weak gravitational deflection angle of massive particles in Brahma-Chen-Yeom rotating black hole. $r_S=r_R=10^8M$, $a=0.5M$, $\lambda=0.1$. We choose a large initial impact parameter since Eq.~\eqref{deltalqg} is a second-order post-Newtonian result for the weak gravitational lensing.}
    \label{fig-4}
\end{figure}

For the unbound particles, we show the deflection angle $\delta$ against the impact parameter $b$ in Fig.~\ref{fig-3} and Fig.~\ref{fig-4}. Specifically, fixing $r_S=r_R=10^8M$ and $v=0.7c$, we plot $\delta$ against $b$ in Fig.~\ref{fig-3} respectively with $\lambda=0$ (Kerr spacetime), $0.1$, $1$, $2$, $5$ and $a=-0.5M$, $0.5M$. 
As we can see, $\delta$ monotonically decreases as $b$ increases for all quantum parameters and spin parameters considered, this agrees with our intuition to gravity that the farther apart the trajectory and the lens, the weaker the deflection effect.
The deflection angle of retrograde orbits (dotted lines) is larger than that of prograde ones (solid lines), this phenomenon is not affected by other parameters, which can be predicted by the third term of Eq.\eqref{deltalqg}.
Fixing $r_S=r_R=10^8M$, $a=0.5M$, and $\lambda=0.1$, we plot $\delta$ against $b$ in Fig.~\ref{fig-4} respectively with $v=0.6c$, $0.7c$, $0.8c$, $0.9c$, $c$ (photon). 
It is shown that the lower the speed of particles, the larger the deflection angle. This is consistent with the conclusion of \cite{liu2016gravitational,pang2019gravitational}, and the intuition that the faster particles will be more difficult to be "dragged" by the black hole for the same impact parameter.

With Eq.~\eqref{deltalqg}, we can set specific values for position of the source $u_S$, position of the receiver $u_R$, the velocity of particles $v$, the spin parameter $a$, and the quantum parameter $\lambda$ to investigate the finite or infinite deflection angle of massless or massive particles for rotating or nonrotating black holes considering or not considering LQG. For example, (a) when $a=0$, $u_S=u_R=0$ and $v=1$, Eq.~\eqref{deltalqg} reduces to the infinite-distance deflection angle of photons in nonrotating black holes in LQG \cite{fu2022gravitational}, (b) when $\lambda=0$, Eq.~\eqref{deltalqg} reduces to the finite-distance deflection angle of massive particles in Kerr spacetime \cite{li2020thefinitedistance}. Moreover, some cases that Eq.~\eqref{deltalqg} covers have never be calculated before.

\subsection{Observational constraint on $\lambda$}
With the photons that come from infinity and graze the Sun, we give an observational constraint on the quantum parameter $\lambda$ on the scale of the solar system.

Let $v=1$ and $u_S=0$, Eq.~\eqref{deltalqg} reduces to
\begin{equation}
    \begin{aligned}
        \delta_{\text{theory}} = & M\frac{2\left(\sqrt{1-b^{2} u_{R}^{2}} +1\right)}{b} -Ma\frac{2\left(\sqrt{1-b^{2} u_{R}^{2}} +1\right)}{b^{2}}\\
            & +M^{2}\Bigg\{\frac{bu_{R}\left[ 16\lambda \left( b^{2} u_{R}^{2} -1\right) -7b^{2} u_{R}^{2} +15\right]}{4b^{2}\sqrt{1-b^{2} u_{R}^{2}}}\\
            & +\frac{(16\lambda -15)\left[\arcsin (bu_{R} )-\pi \right]}{4b^{2}}\Bigg\}\\
            & +\mathcal{O} (M^{3} ,M^{2} a,a^{2} M,a^{3} ),
        \end{aligned}
\end{equation}
which corresponds to the deflection angle of the grazing photons on the assumption that the source is at infinity. For the solar-system experiments, the photon passing the Sun at a distance $d$ is deflected by an angle (Sec.7.1 of \cite{will2018theory})
\begin{equation}
    \delta_{\text{measure}} =\frac{1}{2}(1+\gamma) \frac{4 M_{\odot}}{d} \frac{1+\cos \varphi}{2},
\end{equation}
in which $M_{\odot}$ is the mass of the Sun, $d$ is the distance of the closest approach of the photon, $\varphi$ denotes the angle between the Earth-Sun line and the incoming direction of the photon, $\gamma$ represents the parameter relating the deflection of photons. To be consistent with the notation widely used by others, in and only in this subsection, $\gamma$ represents the deflection parameter. For the grazing photon, $\varphi\approx 0$ and $d$ is approximately equal to the radius of the Sun ($d \approx R_{\odot}$) \cite{will2018theory}. Using $\delta_{\text{theory}}= \delta_{\text{measure}}$, the quantum parameter is expressed in terms of the observables
\begin{equation}
    \begin{aligned}
        \lambda = & \frac{R_{\odot }}{4M_{\odot }\left[ R_{\odot } u_E\sqrt{1-R_{\odot }^{2} u_E^{2}} -\arcsin (R_{\odot } u_E )+\pi \right]} \cdot \\
            & \Bigg\{-2\gamma +\frac{15M_{\odot } \left[ \pi -\arcsin (R_{\odot } u_E ) \right]}{4R_{\odot }} -\frac{2a_\odot}{R_{\odot }} +\\
            & \frac{M_{\odot } u_E\left( 15-7R_{\odot }^{2} u_E^{2}\right)}{4\sqrt{1-R_{\odot }^{2} u_E^{2}}} +\frac{2(R_{\odot } -a_\odot)\sqrt{1-R_{\odot }^{2} u_E^{2}}}{R_{\odot }}\Bigg\},
        \end{aligned}
        \label{quantumparameter}
\end{equation}
in which we have used $M=M_{\odot}$, $b=R_{\odot}$, $a=a_\odot$ and $u_R=u_E$. $a_\odot$ is the spin parameter of the Sun, $u_E$ is the inverse of the distance between the Earth and the Sun.

Various observation data have been used to the measurement of the deflection parameter $\gamma$ \cite{lebach1995measurement,fomalont2009progress,shapiro2004measurement,lambert2009determining,lambert2011improved,froeschle1997determination,treuhaft1991measurement,bolton2006constraint}, here we adopt the result in \cite{lambert2011improved} $\gamma=0.99992 \pm 0.00012$, which is obtained by analyzing more than seven million group delays measured by very long baseline interferometry between August 1979 and August 2010. The angular momentum of the Sun $S_{\odot} = 1.92\times 10^{41} \mathrm{kg \cdot m^2/s}$ is given by Iorio with helioseismology \cite{iorio2012constraining}. With Eq.~\eqref{quantumparameter}, we obtain $-2.9039<\lambda<14.9153$. Considering $\lambda>0$, the constraint on the quantum parameter is finally derived as
\begin{equation}
    0<\lambda<14.9153.
\end{equation}

\section{Conclusion}
\label{sec-5}
Based on LQG, people proposed various schemes to circumvent the classical singularity in cosmological and black hole spacetimes. In this paper, we concentrate on the rotating black hole in LQG constructed by Brahma $et\ al.$ and study the weak gravitational deflection angle of massive particles in such spacetime. 

GW method, a geometric method that can be applied to the investigation of finite-distance deflection angle of timelike or lightlike orbits in SAS spacetimes, is adopted. Different from the treatment in previous literatures, we refine the calculation of GW method and obtain a simplified formula. The double integral in the geometric expression of deflection angle is replaced by a single integral with respect to the azimuthal coordinate. 

To explore the effect of quantum parameter on deflection angle and provide a theoretical expression closer to the real situation, the finite-distance weak gravitational deflection angle of massive particles in Brahma-Chen-Yeom polymerized rotating black hole is obtained. It shows that the quantum parameter appears in the second order and yields a repulsive effect to particles, i.e. the larger the quantum parameter, the smaller the deflection angle. Our result covers the finite-distance or infinite-distance deflection angle of massless or massive particles in rotating or nonrotating black holes with or without quantum corrections. Considering the photons coming from the infinity and grazing the Sun, we express the quantum parameter with observables, and give a constraint on the quantum parameter $0<\lambda<14.9153$ by using the observational results in solar system.

\section*{Acknowledgments}
This work was supported in part by the National Key Research and Development Program of China Grant No. 2021YFC2203001 and in part by the NSFC (No.~11920101003 and No.~12021003). Z. Cao was supported by ``the Interdiscipline Research Funds of Beijing Normal University" and by CAS Project for Young Scientists in Basic Research YSBR-006.

\bibliographystyle{apsrev}
\bibliography{refs}

\appendix
\section{The geodesic curvature of the radial curve}
\label{appendix-1}
For a radial curve in the space equiped with metric~\eqref{metricalpha}, we have $\mathrm{d}\phi=0$ and
\begin{equation}
    \frac{\mathrm{d} r}{\mathrm{d} l} =\frac{1}{\sqrt{\alpha _{rr}}} ,
    \label{appeq1}
\end{equation}
where we have dropped '$( r)$' from '$\alpha _{rr}( r)$' for simplicity. Thus
\begin{equation}
    \frac{\mathrm{d}^{2} r}{\mathrm{d} l^{2}} = \frac{\mathrm{d} r}{\mathrm{d} l} \cdot \frac{\mathrm{d}}{\mathrm{d} r}\left(\frac{\mathrm{d} r}{\mathrm{d} l}\right) =-\frac{\alpha _{rr,r}}{2\alpha _{rr}^{2}}  ,\quad
    \frac{\mathrm{d}^{2} \phi }{\mathrm{d} l^{2}} = 0.
    \label{appeq2}
\end{equation}
What's more
\begin{equation}
\Gamma _{\ \ rr}^{r} =\frac{\alpha _{rr,r}}{2\alpha _{rr}} ,\quad \Gamma _{\ \ rr}^{\phi } =0.
\label{appeq3}
\end{equation}
According to Eqs.~\eqref{appeq1}, \eqref{appeq2} and \eqref{appeq3}, we obtain
\begin{equation}
\frac{\mathrm{d}^{2} r}{\mathrm{d} l^{2}} +\Gamma _{\ \ rr}^{r}\frac{\mathrm{d} r}{\mathrm{d} l}\frac{\mathrm{d} r}{\mathrm{d} l}  =0,\quad
\frac{\mathrm{d}^{2} \phi }{\mathrm{d} l^{2}} +\Gamma _{\ \ rr}^{\phi }\frac{\mathrm{d} r}{\mathrm{d} l}\frac{\mathrm{d} r}{\mathrm{d} l}  =0.
\end{equation}
Namely, an outward radial curve in $M^{(\alpha 2)}$ satisfies the geodesic equation, and its geodesic curvature vanishes.

\end{document}